# Phase slip centers in a two-band superconducting filament: application to MgB$_2$

**V. N. Fenchenko, Y. S. Yerin**

B.Verkin Istitute for Low Temperature Physics and Engineering of the National Academy of Science of Ukraine, 47 Lenin Ave., Kharkov, 61103, Ukraine

E-mail: yerin@ilt.kharkov.ua

Within the framework of non-stationary Ginzburg-Landau equations generalized to the case of two order parameters, we investigated dynamic characteristics of two-band superconducting MgB$_2$ filaments in a voltage-driven regime. We calculated current-voltage characteristics of superconducting MgB$_2$ filaments of different lengths and compared with a single-band superconductor of corresponding sizes. Despite the presence of two interacting bands for small values of the ratio of effective Cooper pairs' weights, phenomenological constants of diffusion and large lengths of channels, the well-known S-form of current-voltage characteristic for the single-band superconductor with fluctuations of the current density in a certain interval of voltage also takes place in the case of the two-band superconductor. An analysis of spectra of the current density and the time-evolution of the order parameters for long channels points out that such unusual forms of current-voltage characteristics are caused by the occurrence of either chaotic oscillations of the moduli of the order parameters or the interference of some of their oscillation modes. A further increase in the ratio of effective weights of the Cooper pairs and in phenomenological constants of diffusion smoothes these oscillations and transforms the system into an ordered state, where the S-form of current-voltage characteristics becomes more flattened. The latter can be a distinctive feature of two-band superconducting long channels.

## 1. Introduction

One of the most intensively studied class of phenomena related to non-equilibrium superconductivity is the formation of the resistive state in long quasi-one dimensional channels and two-dimensional wide films. As is well known, when the transport current exceeds the critical current, the superconducting state does not disappear completely. Efforts to keep superconductivity lead to that at individual points and at certain instants of time the modulus of the order parameter vanishes and its phase undergoes jumps by multiples of $2\pi$. These points, whose size is of order of the coherence length in the 1D superconducting system, and at which the modulus of the order parameter oscillates between zero and its equilibrium (maximum) value, are called phase-slip centers (PSCs) [1].

Experimentally, the occurrence of an inhomogeneous resistive structure and, in particular, PSCs in superconductors is detected by curvatures on current-voltage characteristics (CVCs): regular voltage steps in a current-driven regime and an S-shape form in a voltage-driven regime.

From the theoretical point of view, the simplest description of PSCs is provided by the time-dependent Ginzburg-Landau (TDGL) model which is a generalization of the usual Ginzburg-Landau theory with the inclusion of relaxation processes. The TDGL model is a universally recognized "expert" at investigation into these phenomena, and it often gives a good qualitative and quantitative picture of the superconducting system dynamics near the critical temperature $T_c$.

Until recently. investigations into the non-equilibrium superconductivity were restricted to superconductors with one energy gap with conventional or unconventional pairing mechanism. The discovery of multiband superconductors such as MgB$_2$ [2] and later the family of iron-based superconductors [3] stimulates many theoretical and experimental works in order to exhibit novel effects which have no analogies in single-band superconductors. For instance, it was found both theoretically [4-

7] and experimentally [8,9] that vortices at the surface of the two-band superconductor MgB$_2$ can form highly inhomogeneous lattices, which differ from classical triangular structures in the single-band superconductor. Earlier it was shown [10] that PSCs can be visualized as topological defects of the type of vortices in the 1D space-time that form a periodic structure similar to the Abrikosov lattice in type-2 superconductor. Hence, there it is reasonable to expect some new kind of features in the behavior of the PSCs in the two-band superconducting system. It means that PSCs phenomena must also be re-investigated both theoretically and experimentally.

To begin with, we define a concept of the PSC in two-band superconductors. We will call PSCs such places in the superconducting filament in which *both* the order parameters vanish and their phases slip by $2\pi$. From this definition, some questions arise: Are there any essential differences between the formation and dynamics of PSCs in the single-band superconductor and the two-band analogue? If those differences are present, how are they realized on the CVC? Do the moduli of the order parameters oscillate synchronously? If one order parameter becomes zero in a certain area and its phase slips by $2\pi$, does it mean that the second order parameter undergoes the same process at the same time? To the best of our knowledge, there are no published theoretical works that study the above formulated problems. So, in this paper, we make up this deficiency and investigate PSCs theoretically in a narrow superconducting channel made from a two-band superconductor, namely MgB$_2$, in the voltage-driven regime.

## 2. Basic equations

Our approach is based on numerical solution of non-stationary two-band Ginzburg-Landau equations, which can be written in the following form:

$$\frac{\hbar^2}{2m_1 D_1}\left(\frac{\partial \psi_1}{\partial t} + \frac{2e}{\hbar}i\Phi\psi_1\right) - \frac{\hbar^2}{2m_1}\nabla^2\psi_1 + \alpha_1\psi_1 + \beta_1|\psi_1|^2\psi_1 - \gamma\psi_2 = 0, \quad (1)$$

$$\frac{\hbar^2}{2m_2 D_2}\left(\frac{\partial \psi_2}{\partial t} + \frac{2e}{\hbar}i\Phi\psi_2\right) - \frac{\hbar^2}{2m_2}\nabla^2\psi_2 + \alpha_2\psi_2 + \beta_2|\psi_2|^2\psi_2 - \gamma\psi_1 = 0, \quad (2)$$

$$j = -\frac{\sigma}{c}\nabla\Phi - \frac{ie\hbar}{m_1}\left(\psi_1^*\nabla\psi_1 - \psi_1\nabla\psi_1^*\right) - \frac{ie\hbar}{m_2}\left(\psi_2^*\nabla\psi_2 - \psi_2\nabla\psi_2^*\right). \quad (3)$$

where $\psi_{1,2}$ are the order parameters in the superconducting filament, $\Phi$ is the electric potential, $j$ is the density of the supercurrent flowing through the system, $D_i$ are phenomenological diffusion constants, $\sigma$ is the normal conductivity, $m_i$ are effective masses of the Cooper pairs. The rest of the symbols, $\alpha_i$, $\beta_i$ and $\gamma$, represent phenomenological Ginzburg-Landau coefficients that nevertheless can be calculated from the microscopic theory for the case of the two-band superconductivity. The last parameter $\gamma$ is sp[ecific to the superconductor with two bands and describes the interband coupling. Zhitomirsky, Dao and Silaev, and Babaev in their papers [11,12] provided a microscopic background for $\alpha_i$, $\beta_i$, $\gamma$ and obtained analytical expressions for these Ginzburg-Landau parameters via electron-phonon constants $\lambda_{ij}$, the density of states on the Fermi level $N_i$ and critical temperature $T_c$:

$$\alpha_1 = \left(\frac{\lambda_{22}}{\det\lambda} - \ln\frac{T_c e^{1/\lambda}}{T}\right)N_1, \quad (4)$$

$$\alpha_2 = \left(\frac{\lambda_{11}}{\det\lambda} - \ln\frac{T_c e^{1/\lambda}}{T}\right)N_2, \quad (5)$$

$$\beta_i = \frac{7\zeta(3)N_i}{16\pi^2 T_c^2}, \quad (6)$$

$$\gamma = \frac{\lambda_{12}N_1}{\det\lambda}. \quad (7)$$

Here $\lambda$ is the largest eigenvalue of the matrix $\lambda_{ij}$. Since we consider the vicinity of $T_c$, we can rewrite the expressions for $\alpha_i$ as follows:

$$\alpha_1 = (a_1 - \tau) N_1, \tag{8}$$

$$\alpha_2 = (a_2 - \tau) N_2. \tag{9}$$

Here $a_1 = \dfrac{\lambda_{22}}{\det \lambda} - \dfrac{1}{\lambda}$, $a_2 = \dfrac{\lambda_{11}}{\det \lambda} - \dfrac{1}{\lambda}$, and $\tau = 1 - \dfrac{T}{T_c}$.

Next, for convenience, we introduce the dimensionless parameters: $x = \xi_{10} x'$, $t = \dfrac{\xi_{10}^2}{D_1} t'$, $\psi_i = |\psi_{10}| \psi_i'$,

$\gamma = |\alpha_1(0)| \gamma'$, $j = \dfrac{e \hbar |\psi_{10}|^2}{m_1 \xi_{10}} j'$, $\Phi = \dfrac{\hbar D_1}{2 e \xi_{10}^2} \Phi'$, $\sigma = \dfrac{2 |\psi_{10}|^2 \xi_{10}^2 e^2 c}{m_1 D_1} \sigma'$, where $\xi_{10} = \sqrt{\dfrac{\hbar^2}{2 m_1 |a_1 - 1| N_1}}$ and

$|\psi_{10}| = \sqrt{\dfrac{|a_1 - 1| N_1}{\beta_1}}$ are the coherence length and the equilibrium value of the modulus of the order

parameter, respectively, for the first band without interband coupling at $T = 0$. For clarity, we drop the prime in dimensionless units in the calculations that follow.

Since we assume the transverse dimension of the two-band superconducting filament to be small compared to the coherence lengths and the penetration depths, we can consider the system as 1D and neglect effects due to the magnetic field of the current. Taking into account the continuity equation for the current density, we can reduce Ginzburg-Landau equations (1-3) to the following form:

$$\frac{\partial \psi_1}{\partial t} + i \Phi \psi_1 - \partial_x^2 \psi_1 + \frac{a_1 - \tau}{|a_1 - 1|} \psi_1 + |\psi_1|^2 \psi_1 - \gamma \psi_2 = 0, \tag{10}$$

$$kd \left( \frac{\partial \psi_2}{\partial t} + i \Phi \psi_2 \right) - k \partial_x^2 \psi_2 + \frac{(a_2 - \tau)}{|a_1 - 1|} n \psi_2 + n |\psi_2|^2 \psi_2 - \gamma \psi_1 = 0, \tag{11}$$

$$\partial_x^2 \Phi = -i \left( \psi_1^* \partial_x^2 \psi_1 - \psi_1 \partial_x^2 \psi_1^* \right) - ik \left( \psi_2^* \partial_x^2 \psi_2 - \psi_2 \partial_x^2 \psi_2^* \right). \tag{12}$$

where $k = \dfrac{m_1}{m_2}$, $d = \dfrac{D_1}{D_2}$ and $n = \dfrac{N_2}{N_1}$.

Boundary conditions to equations (10-12) are

$$\psi_i(0, t) = |\psi_i^{(0)}|, \ \psi_i(L, t) = |\psi_i^{(0)}| \exp(iVt), \tag{13}$$

$$\Phi(0, t) = 0, \ \Phi(L, t) = V, \tag{14}$$

where $V$ is the applied voltage at the ends of the filament, $L$ is the filament length, and $|\psi_i^{(0)}|$ are the bulk values of the order parameters modules. As can be found from a numerical solution of the equilibrium system of GL equations,

$$\frac{a_1 - \tau}{|a_1 - 1|} \psi_1^{(0)} + |\psi_1^{(0)}|^2 \psi_1^{(0)} - \gamma \psi_2^{(0)} = 0, \tag{15}$$

$$\frac{(a_2 - \tau)}{|a_1 - 1|} n \psi_2^{(0)} + n |\psi_2^{(0)}|^2 \psi_2^{(0)} - \gamma \psi_1^{(0)} = 0. \tag{16}$$

In order to better understand any possible qualitative and quantitative differences between the two-band superconducting filament and a single-band one, we also solve non-stationary GL equations for the superconductor with single band for parameters related to the first band of the two-band superconductor, e.g. $\alpha = -|a_1 - 1| \tau$ and the phenomenological diffusion coefficient $D = D_1$. In addition, we normalize the modulus of the order parameter by $|\psi_{10}| = \sqrt{\dfrac{|a_1 - 1| N_1}{\beta_1}}$, where $\beta = \beta_1$, and consider a single-band superconductor with critical temperature equal to $T_c$ of the two-band superconductor. Such a choice allows us to use identical time, length, current and voltage units for correct comparison of dynamic characteristics for single-band and two-band superconducting systems. So, taking into account this assumption, we obtain the well-known dimensionless form of single-band non-stationary GL equations:

$$\frac{\partial \psi}{\partial t} + i\Phi\psi - \partial_x^2\psi - \tau\psi + |\psi|^2\psi = 0, \tag{17}$$

$$\partial_x^2\Phi = -i\left(\psi^*\partial_x^2\psi - \psi\partial_x^2\psi^*\right). \tag{18}$$

with boundary conditions

$$\psi(0,t) = |\psi^{(0)}|, \quad \psi(L,t) = |\psi^{(0)}|\exp(iVt), \tag{19}$$

$$\Phi(0,t) = 0, \quad \Phi(L,t) = V. \tag{20}$$

where $|\psi^{(0)}| = \sqrt{\tau}$ is the bulk value of the modulus of the order parameter.

### 3. Numerical results and discussions

We start with the analysis of PSCs in the superconducting filament, focused on MgB$_2$ parameters. It is common knowledge that this two-band superconductor has an electron-phonon pairing mechanism with coupling constants (Coulomb matrix elements are included) $\lambda_{ij} = \begin{pmatrix} 0.807 & 0.118 \\ 0.086 & 0.276 \end{pmatrix}$ and the ratio of the two densities of states $n = 1.372$ [13]. Based on these values, we calculated dimensionless parameters $a_1 = 0.0869$, $a_2 = 2.5847$, $\gamma = 0.6079$. The temperature of the system was supposed to be $T = 0.9T_c$ ($\tau = 0.1$). During the numerical simulations, we varied only two parameters $k$ and $d$, which for the case of dirty superconductor can be interpreted as the ratio of intraband diffusion coefficients due to nonmagnetic impurity scattering (for the case of dirty superconductor) and the ratio of material dependent parameters for each band, respectively, (the material dependent parameter is the ratio of relaxation time of the amplitude of the order parameter to the phase relaxation time). Definitions of $k$ and $d$ permit to change these parameters during our calculations
Solving corresponding equations (10-14) and (17-20), we calculated CVCs for single-band and two-band superconducting channels with different lengths (figure 1).

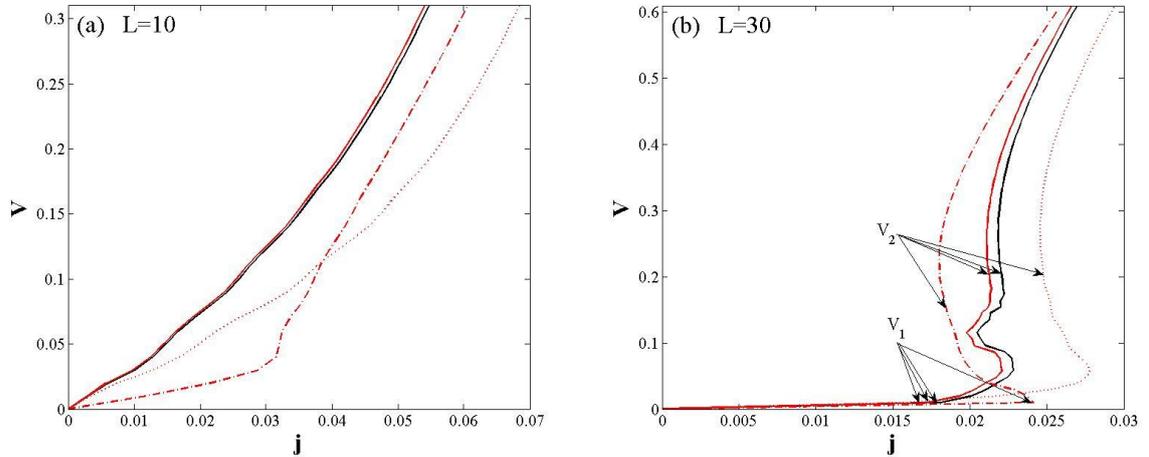

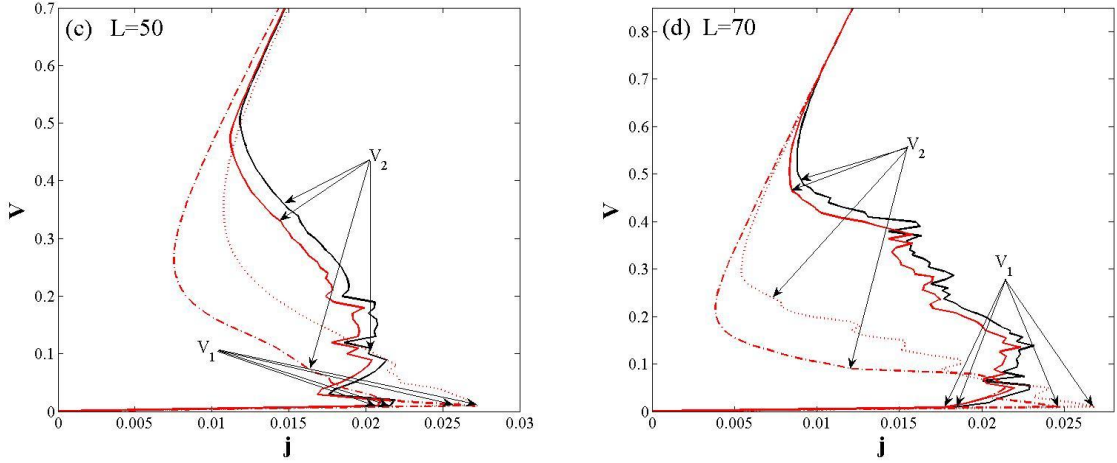

Figure 1 (color online). Current-voltage characteristics of single-band and two-band $MgB_2$ superconducting filaments of different lengths (shown on graphics) and for different phenomenological parameters. Black line corresponds to the single-band superconductor, the red solid line to the two-band superconductor with $k=d=1$, the red dotted line to the two-band superconductor with $k=50$, $d=1$, and the red dash-and-dot line to the two-band superconductor with $k=50$, $d=10$. The arrows indicate voltage intervals $(V_1, V_2)$, where fluctuations of the current density take place.

We averaged function of the current density $j(t)$ in the range of time from 500 to 10000. The start of the time interval for averaging began with 500 in order to avoid the influence on the final result of relaxation processes occurring in the superconducting system at the initial moment of time. During calculations of CVCs the voltage step was equal to 0.005.

We can see that CVCs of single-band and two-band superconductor with $k=d=1$ either practically coincid, when the channels are small ($L=10$), or are situated near each other, when the system has large size ($L=30$, $L=50$ and $L=70$). Moreover, any qualitative features connected with the presence of two order parameters on the CVC of the channel of $MgB_2$ with the set of parameters $k$ and $d$ are not observed.

Such behavior of CVCs of the two-band superconductor is caused by the relation between equilibrium values of the moduli of the order parameters in the present superconductors. For $MgB_2$ the equilibrium value of the modulus of the first order parameter for the given temperature practically coincides (after dimensionless procedure) with the same value for the single-band superconductor and is much larger than the equilibrium value of the modulus of the second order parameter, i.e. $\left|\psi_1^{(0)}\right| \approx \left|\psi^{(0)}\right| \gg \left|\psi_2^{(0)}\right|$. Since the current density is proportional to the expression $j \propto -V + \left(\left|\psi_1^{(0)}\right|^2 + k\left|\psi_2^{(0)}\right|^2\right)$ for the two-band superconductor and proportional to $j \propto -V + \left|\psi^{(0)}\right|^2$ for the single-band one, the contribution of the second order parameter for $k \leq 1$ to the current density is very small. For this reason we obtain a "neighborhood" of CVCs for the single-band and two-band superconductors with $k=d=1$.

The obtained CVCs have strongly pronounced S-shaped forms, except the case of a short channel ($L=10$). Besides, in a certain interval of voltage denoted as $(V_1, V_2)$, fluctuations of the current density take place for long channels. For the single-band superconducting filament in the voltage driven regime similar S-form CVCs were already observed in experiments and predicted theoretically within the framework of non-stationary Ginzburg-Landau equations with the non-zero depairing factor [14]. The authors [14] noticed that unusual fluctuations of CVCs obtained by the numerical solution of the non-stationary Ginzburg-Landau equations can be got only in the case when the depairing factor is much less than 1. We found out that the S-shaped behavior and accompanying fluctuations of the current density obtained in the gapless case for the single-band superconductor when the depairing factor was much larger than 1.

With increasing $k$ and fixed value of $d$, the interval of the voltage where fluctuations of CVCs are realized is narrowed. The increasing contribution to the current density of the second order parameter leads to gradual disappearance of fluctuations of the current density on CVCs and, hence, to the ordering of the system.

If now we fix the value of $k = 1$ and change $d$, then for any value of the ratio of phenomenological constants of diffusion the CVC of the system differs very weakly from the single-band superconducting channel at the qualitative level. For large $d$, with increasing parameter $k$ the interval of fluctuations of the current density becomes smaller, and CVC takes a more flattened form in comparison with the case when $k = 50$ and $d = 1$.

To find out possible reasons for the occurrence of such fluctuations on the CVCs of the long channels, we analyzed the behavior of time dependences of the current density and their spectra for different values of voltage. For clearness, we considered long channels of two-band and single-band superconductors where these oscillations are most expressed, i.e. $L = 70$.

As can be seen from figure 1, there are three basic regimes in the behavior of the current density. For small voltage regime until $V = V_1$ (for each set of parameters $k$ and $d$ as well as for the length of superconducting channel voltage $V_1$ has certain value), where the CVCs do not fluctuate yet, the time dependence of the current density shows sinusoidal dependence, because the corresponding spectrum of the function $j(t)$ (fig. 2 (a)) has a single pronounced peak.

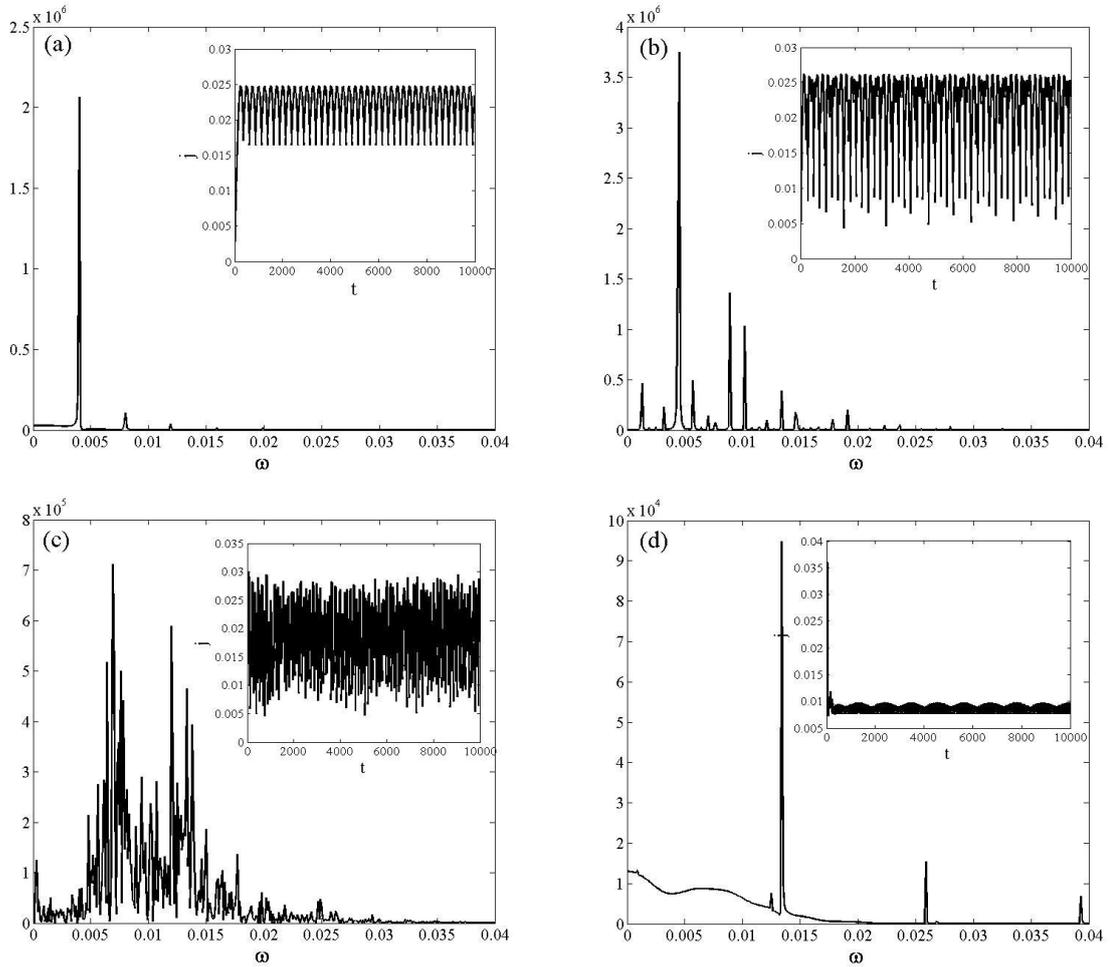

Figure 2. The power spectrum of the time dependence of the current density $j(t)$ for the two-band MgB$_2$ superconducting filament with $L = 70$ and applied voltage (a) $V = 0.05$, (b) $V = 0.1$, (c) $V = 0.25$ and (d) $V = 0.5$. In the insets, $j(t)$ dependences for each voltage value are shown. The time series with values of current density from 0 to 10000 was used for obtaining the spectrum. The time step was equal to 0.01.

Then there is a region $V \in (V_1, V_2)$, where the spectrum of the current density shows a multipeak structure (fig. 2 (b)) at the beginning of this interval, hinting at the origin of the chaos. At voltage values that are in the middle of the present interval, the behavior $j(t)$ becomes chaotic, which specifies the noisy spectrum of the current density-time dependence (fig. 2(c)).

With the growth of voltage, chaotic oscillations and the noisy spectrum gradually disappears, and at $V > V_2$ the third regime takes place in which oscillations of the current density again accept a periodic form (fig. 2 (d)). A further increase in voltage leads to a decrease in the amplitude of these oscillations. There, where the CVC curve becomes a straight line, the current density practically doesn't depend any more on time.

At a qualitative level, similar behavior of the dependence $j(t)$ with three different regimes, one of which has chaotic oscillations, is realized also for the single-band superconducting wire with $L = 70$.

The reason for the occurrence of the three regimes can be understood if we track the evolution of the moduli of the order parameters over the length of a long single-band or two-band superconducting filament. Here we present results of such numerical investigation for a long two-band superconducting channel with $L = 70$ and the parameters $k = d = 1$.

We found out that phases of the order parameters remain coherent and the moduli of the order parameters oscillate synchronously in all the three regions at any places of the filament and at any instants of time. Therefore, in the discussion of PSCs in two-band superconducting channels that follows below, this fact will imply at that both the moduli of the order parameters tend to zero and their phases experience jumps by $2\pi$.

At small voltage values, under the influence of the flowing current the moduli of the order parameters slowly subside and then start to form one oscillating PSC in the center of the filament, which after that changes into 2 PSC in the centers of halves of the channel as voltage is increased. Such behavior is observed for voltage that does not exceed $V_1$.

After the voltage has exceeded $V_1$, more than 2 PSCs start to form in the system. For values of the voltage corresponding to the beginning of the interval $(V_1, V_2)$, 3 PSC are born in the channel. They appear in a periodic manner in the center of the channel and in the centers of its halves (figure 3 (a,b)).

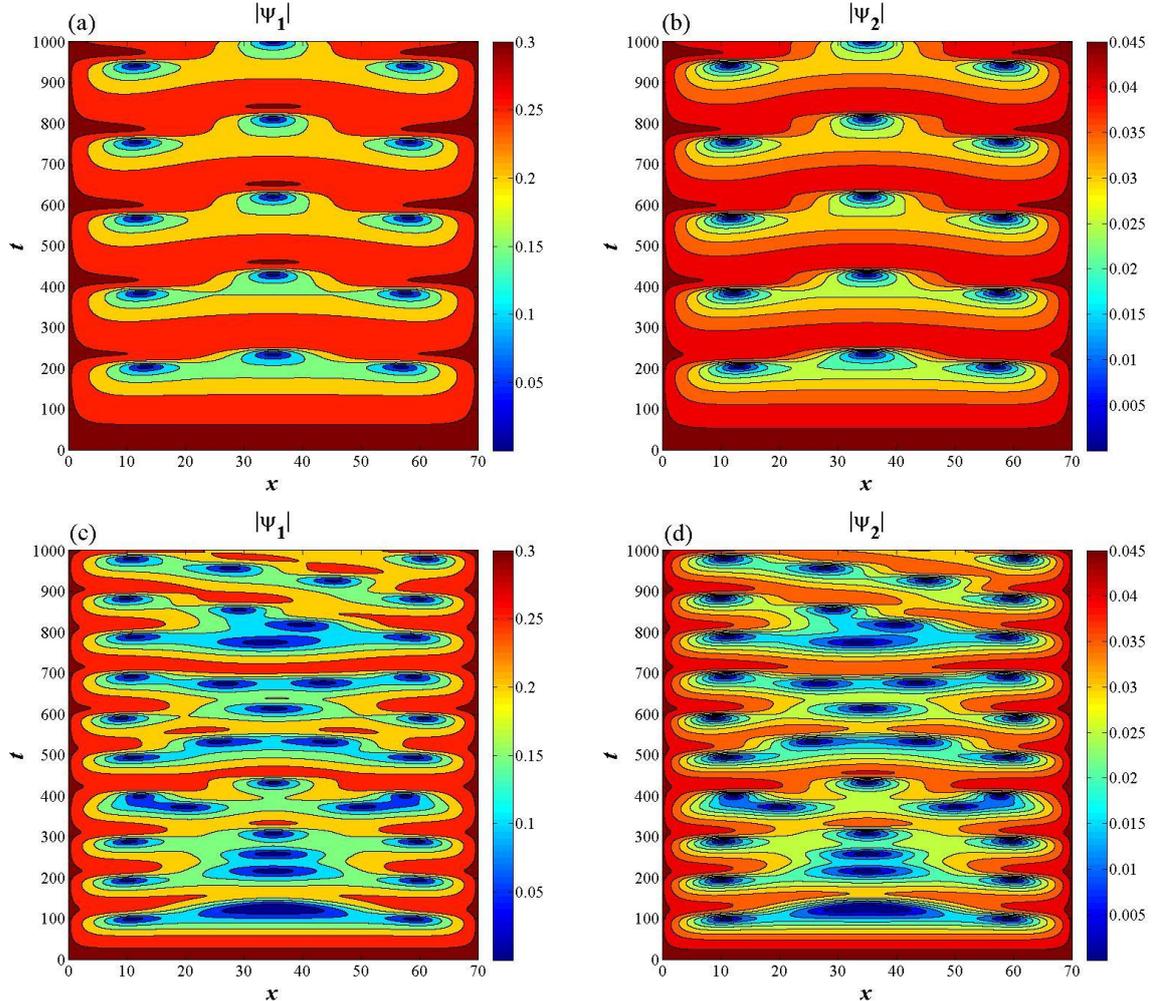

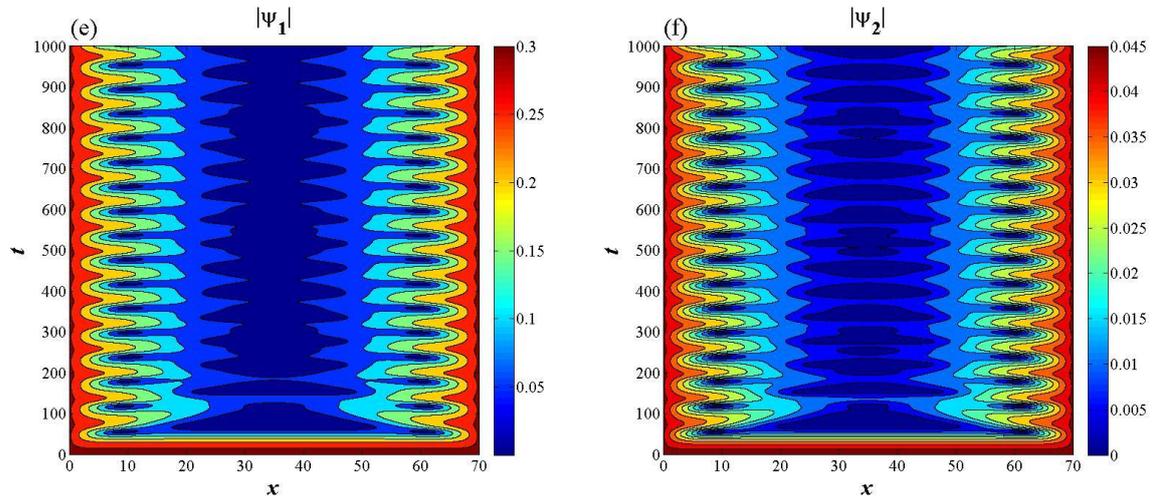

Figure 3 (color online). The time evolution of the moduli of the order parameters in the two-band MgB$_2$ superconducting filament with $L=70$ and applied voltage $V=0.1$ (a, b), $V=0.25$ (c, d) and $V=0.5$ (e, f). The time step was 0.01. The mesh of the spatial grid was taken 0.5.

When voltage values attain the middle of the interval $(V_1, V_2)$, oscillations of order parameters modules have chaotic character (fig. 3 (c,d)). The number of formed PSCs exceeds 3, and the formation of PSCs is random. PSCs form not only in the middle of the channel and in the centers of its halves but also in other places. Further growth of the voltage leads to gradual disappearance of chaotic oscillations of the moduli of the order parameters, and the system returns to the situation with 3 PSC.

Based on the obtained data on evolution of order parameter modules, we can make an important conclusion: the chaos/multipeak structure of the $j(t)$ dependence and, as consequence, fluctuations on CVCs of long channels is caused by chaotic oscillations/interference of several oscillation modes of order parameters modules. At the same time, despite arising chaos or the interference of oscillation modes, the the phases of the order parameters oscillate in a synchronous manner. The last statement doesn't contradict predictions made in [15, 16] concerning the existence of phase soliton structures in two-band superconductors. Firstly, here we have investigated the dynamics of the ground state of two-band superconducting 1D system. Secondly, recently, it has been proved that such soliton states are thermodynamically metastable [17].

At voltage higher than $V_2$, with voltage growth the central PSC gradually turns to a plateau at whose edges the order parameter modules are strongly suppressed and are equal to zero in the central region (fig. 3 (e,f)). This situation corresponds to the approach of CVCs to the straight line.

According to our numerical simulations, chaotic formation of multiple PSCs for single-band or two-band superconducting systems in the second regime continues for channels with lengths no less than $L^{(chaos)} \approx 69$. Reduction of the length up to $L^{(3)} \approx 50$ leads to 3 PSCs near centers of halves of the filament and in its center. From $L^{(3)}$ to $L^{(2)} \approx 37$ one can observe 2 PSCs, which with decreasing of the length begin to shift from the centers of channel halves to the center of the filament. Such movement towards one another induces creation of extra PSC in the center of the channel with length $L > L^{(1)} \approx 25$. When the length of the filament is less than $L^{(1)}$ these 3 PSCs merge among themselves and form a single PSC at the center of the filament. The existence of a single PSC means that there is no interference of different oscillating modes of order parameters modules so the CVC of the system with $L < L^{(1)}$ has no fluctuations.

For the single-band superconducting channel with $L = 70$, the evolution of the order parameter module at qualitative and quantitative levels almost coincides with the above described picture.

However, a different picture arises if we consider two-band superconducting channel with $L = 70$ and the parameters $k \neq 1$ and $d \neq 1$. As it has been discussed above, the increase in the ratio of the effective weights of the Cooper pairs and in the ratio of phenomenological diffusion coefficients is accompanied by smoothing fluctuations of the current density on CVC and ordering of the system To prove this fact, we have calculated the spectrum of the dependence $j(t)$ and tracked again the evolution of the order parameters modules for the two-band superconducting filament with the parameters $k = 50$, $d = 1$ and

$k = 50$, $d = 10$ and for the voltage taken from the middle of the interval $(V_1, V_2)$ corresponding to the second regime (fig. 4).

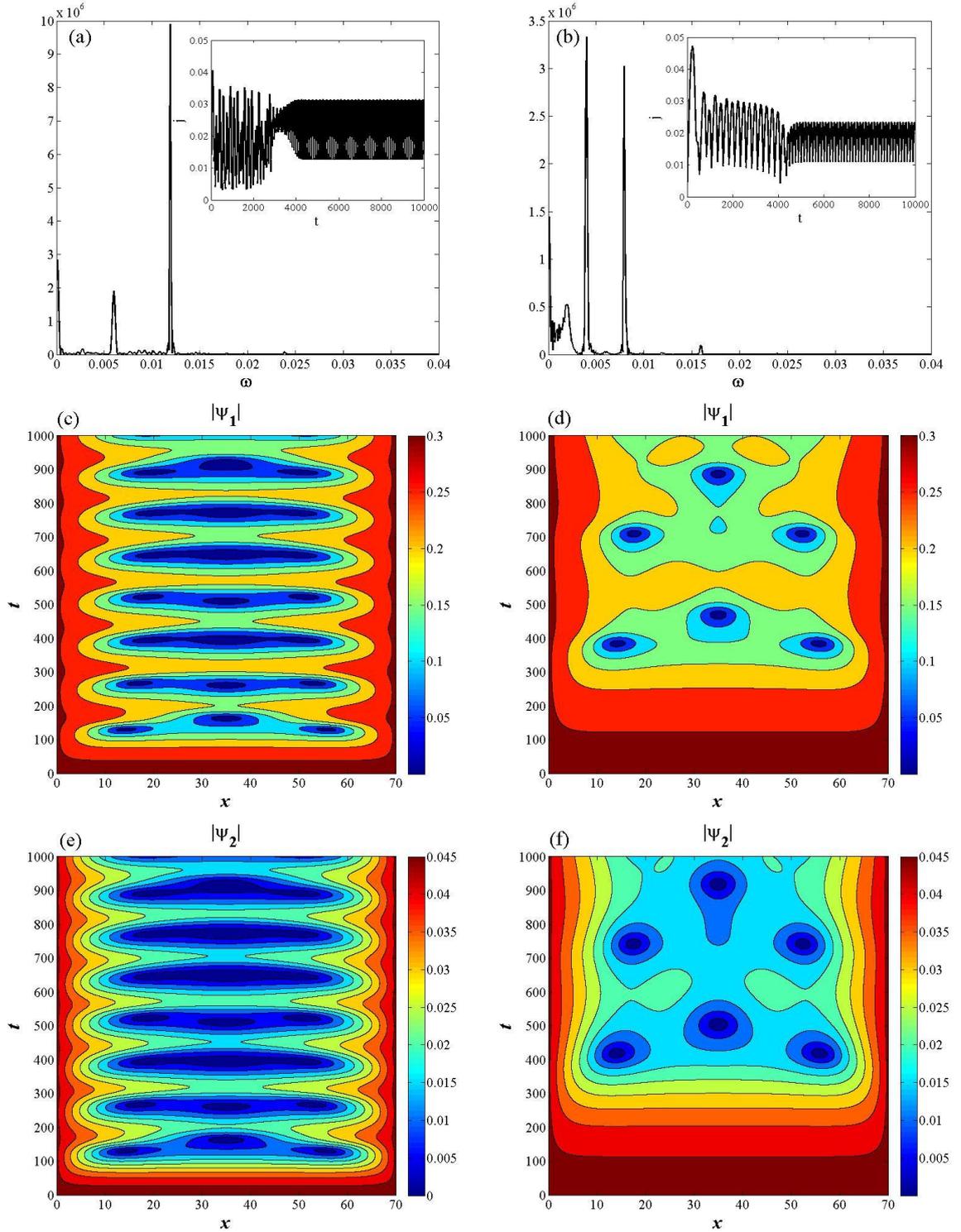

Figure 4. The power spectrum (a,b) and the time evolution of the order parameters modules (c-f) of two-band MgB$_2$ superconducting filament with $L = 70$ for $V = 0.12$, $k = 50$, $d = 1$ (left column) and $V = 0.05$, $k = 50$, $d = 10$ (right column). In the insets, the $j(t)$ dependences are shown.

The shape of the spectrum $j(t)$ and the picture of the evolution of the order parameters modules is clear evidence of the absence of chaos in given systems. In turn, the presence of several peaks on the spectrum of time dependence of the current density allows us to conclude that the observed fluctuations on CVC are caused only by the interference of oscillation modes of the order parameters modules.

## 4. Conclusion

We have shown that, contrary to the expectations, a qualitative picture of the resistive state formation for the two-band superconductor like $MgB_2$ remains the same as in single-band superconducting system.

We have tracked the evolution of order parameters in two-band superconducting $MgB_2$ filaments and shown that for arbitrary voltage the phases of the order parameters remain coherent and the moduli of the order parameters oscillate synchronously at any places of the system and at any instants of time. From this, we have found out that PSCs in the two-band superconductor (at least, $MgB_2$) are points where both the moduli of the order parameters vanish and their phases change jumpwise by multiples of $2\pi$.

We reveal that the CVCs of $MgB_2$ with $k = d = 1$ don't have any structural features and don't differ from similar characteristics of single-band superconductors of the same length. Based on the analysis of data for the evolution of the order parameters, we have established that structure of the CVCs of single-band and two-band superconducting $MgB_2$ filaments consists of three regimes.

The first regime corresponds to periodic oscillations of the moduli of the order parameters with formation of 1 and then 2 PSCs (if the channel is long enough) in systems. The second regime represents processes of initiation, realization and gradual disappearance of chaotic oscillations (for $L > L^{(chaos)}$) or the interference of several oscillation modes of the moduli of the order parameters (for $L^{(1)} < L < L^{(chaos)}$) leading to fluctuations of CVCs. Finally, the third regime represents strong suppression of the moduli of the order parameters and the growth of the area of central PSC. This moment corresponds to the approach of CVCs to a straight line.

With an increase in the ratio of effective weights of the Cooper pairs and in phenomenological constants of diffusion, oscillations of the moduli of the order parameters become ordered in the second regime. As a result, current density fluctuations on CVCs gradually disappear and the CVCs become more flattened. This fact can be considered as a distinctive feature of long two-band superconducting $MgB_2$ channels.

In future, we plan to carry out a more detailed investigation into the chaotic behavior of the current density-time dependence and calculate its Lyapunov exponent for this two-band superconducting system.

## Acknowledgment

This work has been supported by NASU Young Scientists Grant 16-2011. We acknowledge helpful discussions with A.N. Omelyanchouk.